\documentclass[aps,prl,twocolumn,showpacs,superscriptaddress,groupedaddress]{revtex4}  

\usepackage{graphicx}  
\usepackage{dcolumn}   
\usepackage{bm}        
\usepackage{amssymb}   
\usepackage{xcolor}    

\begin{document}

\title{Extraction of beam-spin asymmetries from the hard exclusive $\pi^{+}$ channel off protons in a wide range of kinematics}

\newcommand*{\ANL}{Argonne National Laboratory, Argonne, Illinois 60439}
\newcommand*{\ANLindex}{1}
\affiliation{\ANL}
\newcommand*{\CSUDH}{California State University, Dominguez Hills, Carson, CA 90747}
\newcommand*{\CSUDHindex}{2}
\affiliation{\CSUDH}
\newcommand*{\CANISIUS}{Canisius College, Buffalo, NY}
\newcommand*{\CANISIUSindex}{3}
\affiliation{\CANISIUS}
\newcommand*{\CMU}{Carnegie Mellon University, Pittsburgh, Pennsylvania 15213}
\newcommand*{\CMUindex}{4}
\affiliation{\CMU}
\newcommand*{\SACLAY}{IRFU, CEA, Universit\'{e} Paris-Saclay, F-91191 Gif-sur-Yvette, France}
\newcommand*{\SACLAYindex}{5}
\affiliation{\SACLAY}
\newcommand*{\CNU}{Christopher Newport University, Newport News, Virginia 23606}
\newcommand*{\CNUindex}{6}
\affiliation{\CNU}
\newcommand*{\UCONN}{University of Connecticut, Storrs, Connecticut 06269}
\newcommand*{\UCONNindex}{7}
\affiliation{\UCONN}
\newcommand*{\DUKE}{Duke University, Durham, North Carolina 27708-0305}
\newcommand*{\DUKEindex}{8}
\affiliation{\DUKE}
\newcommand*{\DUQUESNE}{Duquesne University, 600 Forbes Avenue, Pittsburgh, PA 15282 }
\newcommand*{\DUQUESNEindex}{9}
\affiliation{\DUQUESNE}
\newcommand*{\FAU}{Fairfield University, Fairfield CT 06824}
\newcommand*{\FAUindex}{10}
\affiliation{\FAU}
\newcommand*{\FERRARAU}{Universita' di Ferrara , 44121 Ferrara, Italy}
\newcommand*{\FERRARAUindex}{11}
\affiliation{\FERRARAU}
\newcommand*{\FIU}{Florida International University, Miami, Florida 33199}
\newcommand*{\FIUindex}{12}
\affiliation{\FIU}
\newcommand*{\FSU}{Florida State University, Tallahassee, Florida 32306}
\newcommand*{\FSUindex}{13}
\affiliation{\FSU}
\newcommand*{\GWUI}{The George Washington University, Washington, DC 20052}
\newcommand*{\GWUIindex}{14}
\affiliation{\GWUI}
\newcommand*{\ISU}{Idaho State University, Pocatello, Idaho 83209}
\newcommand*{\ISUindex}{15}
\affiliation{\ISU}
\newcommand*{\INFNFE}{INFN, Sezione di Ferrara, 44100 Ferrara, Italy}
\newcommand*{\INFNFEindex}{16}
\affiliation{\INFNFE}
\newcommand*{\INFNFR}{INFN, Laboratori Nazionali di Frascati, 00044 Frascati, Italy}
\newcommand*{\INFNFRindex}{17}
\affiliation{\INFNFR}
\newcommand*{\INFNGE}{INFN, Sezione di Genova, 16146 Genova, Italy}
\newcommand*{\INFNGEindex}{18}
\affiliation{\INFNGE}
\newcommand*{\INFNRO}{INFN, Sezione di Roma Tor Vergata, 00133 Rome, Italy}
\newcommand*{\INFNROindex}{19}
\affiliation{\INFNRO}
\newcommand*{\INFNTUR}{INFN, Sezione di Torino, 10125 Torino, Italy}
\newcommand*{\INFNTURindex}{20}
\affiliation{\INFNTUR}
\newcommand*{\INFNPAV}{INFN, Sezione di Pavia, 27100 Pavia, Italy}
\newcommand*{\INFNPAVindex}{21}
\affiliation{\INFNPAV}
\newcommand*{\ORSAY}{Universit'{e} Paris-Saclay, CNRS/IN2P3, IJCLab, 91405 Orsay, France}
\newcommand*{\ORSAYindex}{22}
\affiliation{\ORSAY}
\newcommand*{\JMU}{James Madison University, Harrisonburg, Virginia 22807}
\newcommand*{\JMUindex}{23}
\affiliation{\JMU}
\newcommand*{\JLU}{Justus Liebig University Giessen, 35392 Giessen, Germany}
\newcommand*{\JLUindex}{24}
\affiliation{\JLU}
\newcommand*{\KNU}{Kyungpook National University, Daegu 41566, Republic of Korea}
\newcommand*{\KNUindex}{25}
\affiliation{\KNU}
\newcommand*{\LAM}{Lamar University, Beaumont, Texas 77705}
\newcommand*{\LAMindex}{25}
\affiliation{\LAM}
\newcommand*{\MISS}{Mississippi State University, Mississippi State, MS 39762-5167}
\newcommand*{\MISSindex}{27}
\affiliation{\MISS}
\newcommand*{\ITEP}{National Research Centre Kurchatov Institute - ITEP, Moscow, 117259, Russia}
\newcommand*{\ITEPindex}{28}
\affiliation{\ITEP}
\newcommand*{\UNH}{University of New Hampshire, Durham, New Hampshire 03824-3568}
\newcommand*{\UNHindex}{29}
\affiliation{\UNH}
\newcommand*{\NRCKI}{Norfolk State University, Norfolk, Virginia 23504}
\newcommand*{\NRCKIindex}{30}
\affiliation{\NRCKI}
\newcommand*{\NSU}{National Research Centre Kurchatov Institute, Petersburg Nuclear Physics Institute, RU-188300 Gatchina, Russia}
\newcommand*{\NSUindex}{31}
\affiliation{\NSU}
\newcommand*{\OHIOU}{Ohio University, Athens, Ohio  45701}
\newcommand*{\OHIOUindex}{32}
\affiliation{\OHIOU}
\newcommand*{\ODU}{Old Dominion University, Norfolk, Virginia 23529}
\newcommand*{\ODUindex}{33}
\affiliation{\ODU}
\newcommand*{\RPI}{Rensselaer Polytechnic Institute, Troy, New York 12180-3590}
\newcommand*{\RPIindex}{34}
\affiliation{\RPI}
\newcommand*{\URICH}{University of Richmond, Richmond, Virginia 23173}
\newcommand*{\URICHindex}{35}
\affiliation{\URICH}
\newcommand*{\ROMAII}{Universita' di Roma Tor Vergata, 00133 Rome Italy}
\newcommand*{\ROMAIIindex}{36}
\affiliation{\ROMAII}
\newcommand*{\MSU}{Skobeltsyn Institute of Nuclear Physics, Lomonosov Moscow State University, 119234 Moscow, Russia}
\newcommand*{\MSUindex}{37}
\affiliation{\MSU}
\newcommand*{\SCAROLINA}{University of South Carolina, Columbia, South Carolina 29208}
\newcommand*{\SCAROLINAindex}{38}
\affiliation{\SCAROLINA}
\newcommand*{\TEMPLE}{Temple University,  Philadelphia, PA 19122 }
\newcommand*{\TEMPLEindex}{39}
\affiliation{\TEMPLE}
\newcommand*{\JLAB}{Thomas Jefferson National Accelerator Facility, Newport News, Virginia 23606}
\newcommand*{\JLABindex}{40}
\affiliation{\JLAB}
\newcommand*{\UTFSM}{Universidad T\'{e}cnica Federico Santa Mar\'{i}a, Casilla 110-V Valpara\'{i}so, Chile}
\newcommand*{\UTFSMindex}{41}
\affiliation{\UTFSM}
\newcommand*{\INSUBRIA}{Universit\`{a} degli Studi dell'Insubria, 22100 Como, Italy}
\newcommand*{\INSUBRIAindex}{42}
\affiliation{\INSUBRIA}
\newcommand*{\BRESCIA}{Universit\`{a} degli Studi di Brescia, 25123 Brescia, Italy}
\newcommand*{\BRESCIAindex}{43}
\affiliation{\BRESCIA}
\newcommand*{\GLASGOW}{University of Glasgow, Glasgow G12 8QQ, United Kingdom}
\newcommand*{\GLASGOWindex}{44}
\affiliation{\GLASGOW}
\newcommand*{\YORK}{University of York, York YO10 5DD, United Kingdom}
\newcommand*{\YORKindex}{45}
\affiliation{\YORK}
\newcommand*{\VT}{Virginia Tech, Blacksburg, Virginia   24061-0435}
\newcommand*{\VTindex}{46}
\affiliation{\VT}
\newcommand*{\VIRGINIA}{University of Virginia, Charlottesville, Virginia 22901}
\newcommand*{\VIRGINIAindex}{47}
\affiliation{\VIRGINIA}
\newcommand*{\WM}{College of William and Mary, Williamsburg, Virginia 23187-8795}
\newcommand*{\WMindex}{48}
\affiliation{\WM}
\newcommand*{\Wupp}{Fachbereich Physik, Universit¨at Wuppertal, D-42097 Wuppertal, Germany}
\newcommand*{\Wuppindex}{49}
\affiliation{\Wupp}
\newcommand*{\YEREVAN}{Yerevan Physics Institute, 375036 Yerevan, Armenia}
\newcommand*{\YEREVANindex}{50}
\affiliation{\YEREVAN}

\newcommand*{\NOWISU}{Idaho State University, Pocatello, Idaho 83209}
\newcommand*{\NOWBRESCIA}{Universit\`{a} degli Studi di Brescia, 25123 Brescia, Italy}

\author{S.~Diehl}
\affiliation{\UCONN}  
\affiliation{\JLU}  
\author {K.~Joo} 
\affiliation{\UCONN}
\author{A.~Kim}
\affiliation{\UCONN}
\author {H.~Avakian} 
\affiliation{\JLAB}
\author{P.~Kroll}
\affiliation{\Wupp}
\author{K.~Park}
\affiliation{\KNU} 
\author {D.~Riser } 
\affiliation{\UCONN} 
\author{K.~Semenov-Tian-Shansky}
\affiliation{\NRCKI} 
\author{K.~Tezgin}
\affiliation{\UCONN} 
\author {K.P.~Adhikari} 
\affiliation{\ODU}
% ...

\collaboration{The CLAS Collaboration}
\noaffiliation

\begin{abstract}
We have measured beam-spin asymmetries to extract the $\sin\phi$ moment $A_{LU}^{\sin\phi}$ from the hard exclusive $\vec{e} p \to e^\prime n \pi^+$ reaction above the resonance region, for the first time with nearly full coverage from forward to backward angles in the center-of-mass. The $A_{LU}^{\sin\phi}$ moment has been measured up to 6.6~GeV$^{2}$ in $-t$, covering the kinematic regimes of Generalized Parton Distributions (GPD) and baryon-to-meson Transition Distribution Amplitudes (TDA) at the same time. The experimental results in very forward kinematics demonstrate the sensitivity to chiral-odd and chiral-even GPDs. In very backward kinematics where the TDA framework is applicable, we found $A_{LU}^{\sin\phi}$ to be negative, while a sign change was observed near 90$^\circ$ in the center-of-mass. The unique results presented in this paper will provide critical constraints to establish reaction mechanisms that can help to further develop the GPD and TDA frameworks.
\end{abstract}

\maketitle


Hard exclusive pseudoscalar meson electroproduction processes offer a unique opportunity to study the structure of the nucleon. They allow one to vary the size of both the probe (i.e. the photon virtuality $Q^2$) and the target (the four-momentum transfer to the nucleon (meson) $t$ ($u$)). These reactions reveal rich information about the structure of the nucleon and the reaction dynamics.  

\begin{figure}[h!]
\begin{center}
    \includegraphics[width=0.24\textwidth]{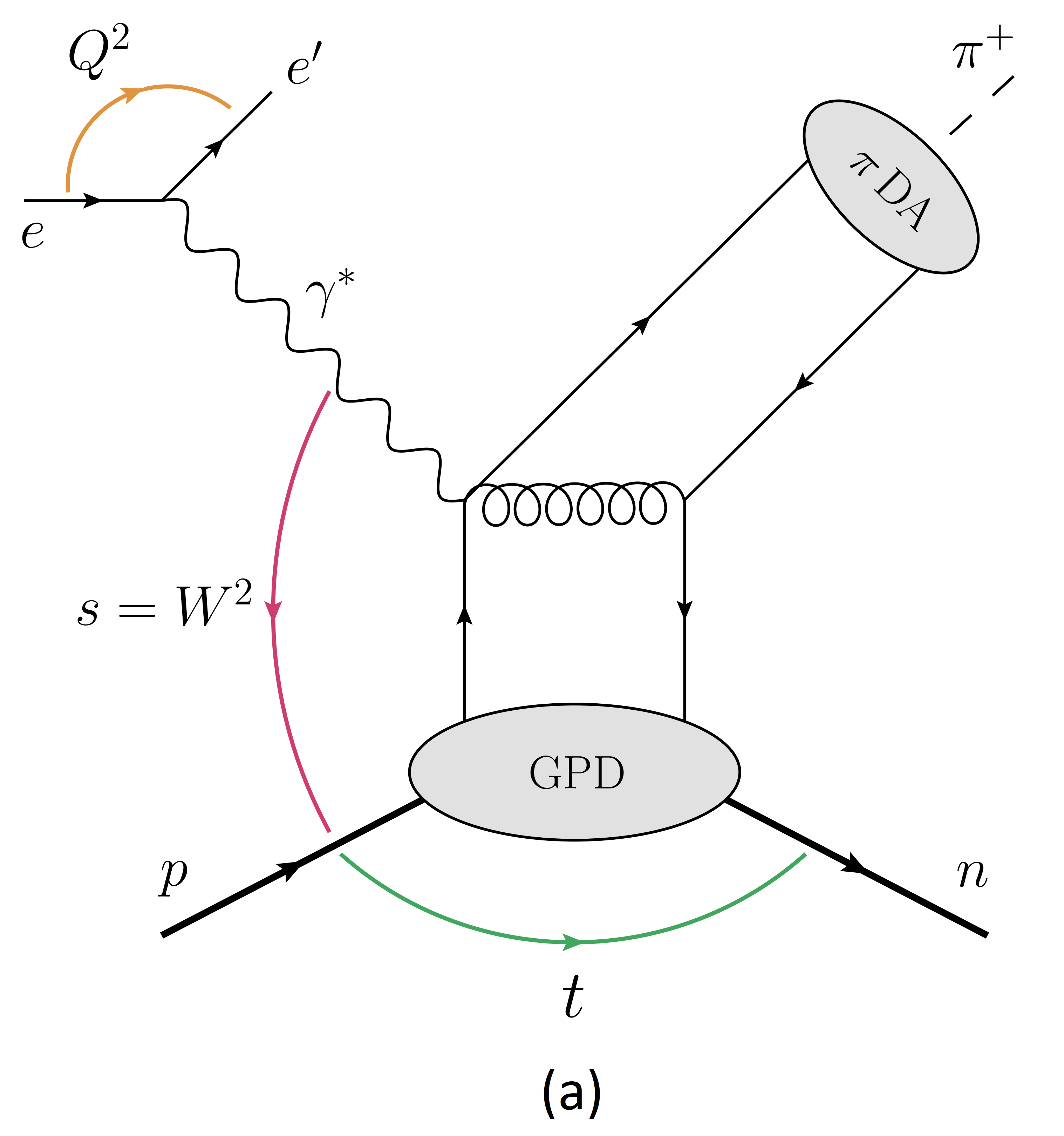}
    \includegraphics[width=0.24\textwidth]{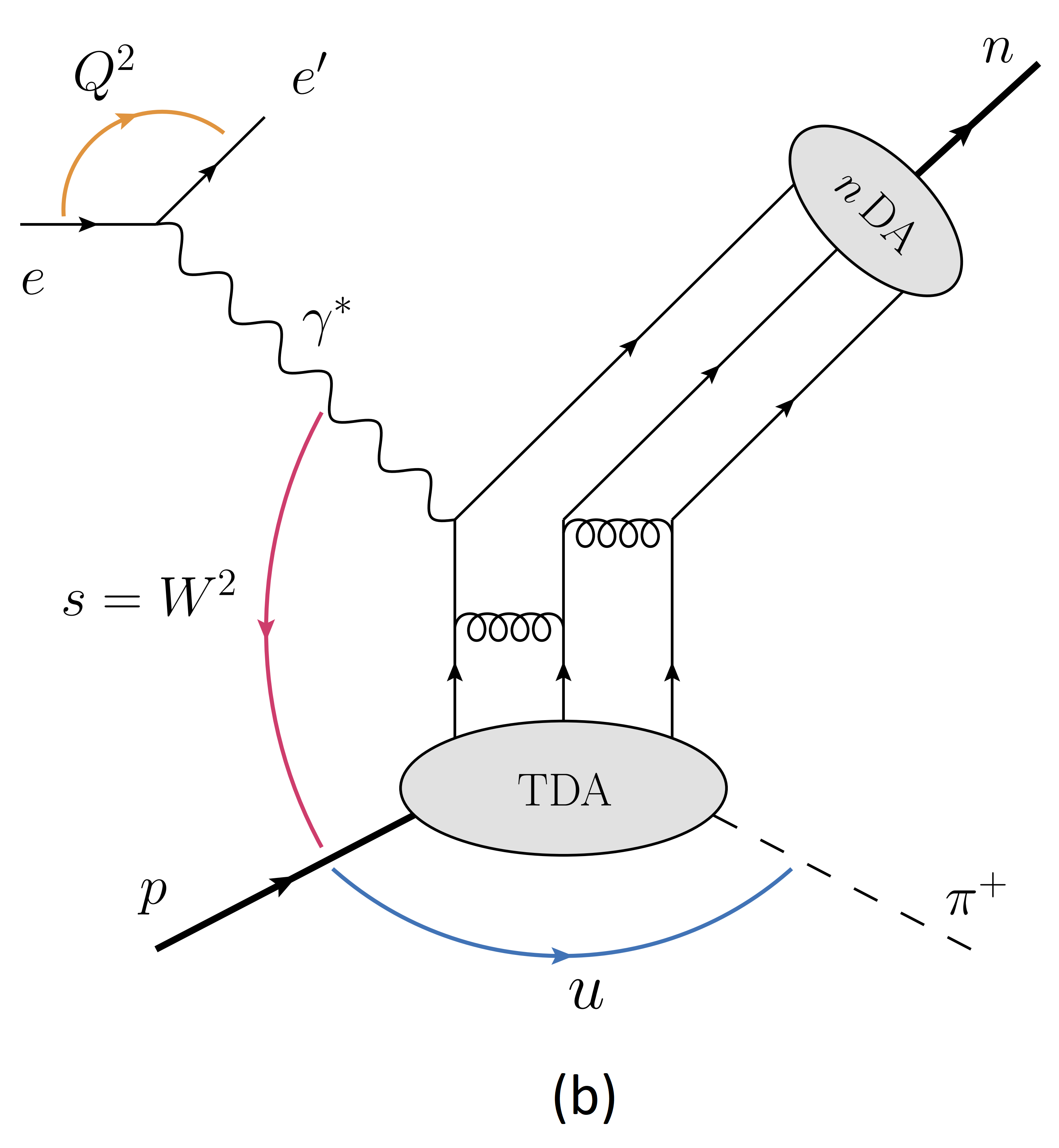}    
	\caption{(a) Exclusive electroproduction of a pion on the proton in very forward kinematics ($-t/Q^2 \ll 1$), described by GPDs~\cite{previous1, previous2}. (b) Factorization of the same reaction in very backward kinematics ($-u/Q^2 \ll 1$), described by TDAs~\cite{PRD85, KPark2018}.}
	\label{fig:production_mechanism}
	\end{center}
\end{figure}

At very forward kinematics ($-t/Q^2~\ll~1$) where the Bjorken limit is applicable, hard exclusive pseudoscalar meson electroproduction can be factorized into a perturbatively calculable hard sub-process at the quark level, $\gamma^{*} q \rightarrow \pi q$, and the hadronic matrix elements which are expressed via the leading twist Generalized Parton Distributions (GPDs) of the nucleon and the pion Distribution Amplitude (DA) \cite{CFS97, previous1, previous2} as shown in Fig.~\ref{fig:production_mechanism} (a). GPDs are light-cone matrix elements that can be expressed as non-local bilinear quark and gluon operators that describe the transition from the initial to the final nucleon and reveal the 3-dimensional structure of the nucleon at the parton level by correlating the internal transverse position of the partons to their longitudinal momentum ~\cite{Mueller:1998fv,Rady96a,Ji97a}. A first experimental confirmation of the applicability of the leading twist GPD framework was provided by deeply virtual Compton scattering (DVCS) experiments at Jefferson Lab (JLab), DESY and CERN~(see, e.g., \cite{CLAS_DVCS:2008, Hermes_DVCS:2009, DVCS5, H1:2009, ZEUS:2009, COMPASS_DVCS:2019}). 

While the DVCS process gives access to all chiral-even GPDs $H$, $\widetilde{H}$, $E$ and $\widetilde{E}$, pseudoscalar meson production is especially helpful in probing the polarized GPDs ($\widetilde{H}$ and $\widetilde{E}$), which contain information about the spatial distribution of the quark spin \cite{Goe01, Jak96}. However, extensive experimental \cite{HERMES02, DeMasi2008, HERMES08, HERMES10, Bedlinskiy2012, Bedlinskiy2014, Kim2017, Bosted_pi02017, Bosted_piplus2017, Bedlinskiy2017, Zhao2019, hallA_2012, hallA_2016, hallA_2017, COMPASS20} and theoretical \cite{previous1, previous2, DK07, DMP17, SS19} investigations of hard exclusive pseudoscalar meson electroproduction in recent years have shown that the asymptotic leading-twist approximation is not readily applicable in the range of kinematics accessible to current experiments. In fact, there are strong contributions from transversely polarized virtual photons that are asymptotically suppressed by $1/Q^{2}$ in the cross sections and have to be considered by introducing chiral-odd GPDs ($H_{T}$, $\widetilde{H}_{T}$, $E_{T}$, and $\widetilde{E}_{T}$) into the framework. For instance for $\pi^0$ and $\eta$ electroproduction, the contributions from transversely polarized virtual photons are significant and the introduction of chiral-odd GPDs is needed to reproduce the measured cross sections as well as large beam- and target-spin asymmetries with GPD models \cite{Bedlinskiy2014, Bedlinskiy2017, Kim2017, Zhao2019, previous1, previous2, Ahmad2009, Goldstein2011}.

A further generalization of the GPD concept has been introduced for non-diagonal transitions where the initial and final states are hadronic states of different baryon number \cite{Frankfurt:1999fp, Pire:2005ax,Pire:2005prd,JLansberg}. In very backward kinematics ($-u/Q^2~\ll~1$) the collinear factorized description can be applied in terms of a convolution of a hard part calculable in perturbative QCD, and the soft parts expressed in terms of nucleon-to-pion baryonic Transition Distribution Amplitudes (TDAs) and the nucleon DA as shown in Fig.~\ref{fig:production_mechanism} (b). Like GPDs, nucleon-to-meson TDAs are defined through hadronic matrix elements of non-local  three-quark light-cone operators. Nucleon-to-meson TDAs are universal functions that parameterize the non-perturbative structure of hadrons.
Within the reaction mechanism involving TDAs, the three-quark core of the target nucleon absorbs most of the virtual photon momentum and recoils forward, while a pion from the mesonic cloud of the nucleon remains with a low momentum heading backward. Therefore, the process brings a bulk of new information on hadronic structure and can be used {\it e.g.} to probe the non-minimal Fock components of hadronic light-cone wave functions. In contrast to the very forward kinematic regime in the Bjorken limit, the contribution of the transversely polarized virtual photon exchange is expected to dominate the process to leading twist-$3$ accuracy in very backward kinematics.
Recent publications on exclusive $\pi^+$ electroproduction by the CLAS collaboration \cite{KPark2018} and on $\omega$ electroproduction from JLab Hall C \cite{Li2017} in very backward kinematics have shown a first indication of the applicability of the TDA model to predict the magnitude and the scaling behavior of the cross section, as well as the dominance of the transverse over the longitudinal cross section at sufficiently large $Q^{2}$ in the backward regime.

The GPD and TDA approaches describe complementary kinematic domains. While GPDs are applicable for small $-t$, TDAs can be applied for small $-u$, corresponding to large $-t$. Although these two approaches deal with domains that are well distinct at asymptotic energies, they are not well separated in the kinematic range accessible to current experiments. Therefore, it is important to investigate in detail the phenomenological differences of the two approaches over a large range of momentum transfer $t$. In previous publications, e.g. \cite{Bosted_piplus2017, KPark2018}, only very limited kinematic regions covering either the GPD or the TDA regime exclusively have been investigated. In this letter, we present a measurement of the beam-spin asymmetries (BSA) for the hard exclusive electroproduction $ep \to e'n\pi^+$ for full $\pi^+$ center-of-mass (CM) angular coverage with a large range of $t$ or $u$.

GPDs and TDAs can be accessed through different observables in exclusive meson production, such as differential cross sections and beam and target polarization asymmetries \cite{Dre1992, Die2005}. The focus of this work is on the extraction of the $A_{LU}^{\sin\phi}$ moment from the beam-spin asymmetry. The beam-spin asymmetry in the one-photon exchange approximation is defined as follows \cite{Dre1992}:
\begin{eqnarray}\label{eq:BSA}
	BSA(t, \phi, x_{B}, Q^{2})  = \frac{d\sigma^{+} - d\sigma^{-}}{d\sigma^{+} + d\sigma^{-}} \nonumber \\ 
	= \frac{A_{LU}^{\sin\phi} \sin\phi}{1 + A_{UU}^{\cos\phi} \cos\phi + A_{UU}^{\cos2\phi} \cos2\phi},
\end{eqnarray}
where $d\sigma^{\pm}$ is the differential cross section for each beam helicity state ($\pm$). For the positive / negative helicity the spin is parallel / anti-parallel to the beam direction. The subscripts $ij$ represent the longitudinal (L) or unpolarized (U) state of the beam and the target, respectively. $\phi$ is the azimuthal angle between the electron scattering plane and the hadronic reaction plane.

Due to the interference between the amplitudes for longitudinal ($\gamma^{*}_{L}$) and transverse ($\gamma^{*}_{T}$) virtual photon polarizations, 
the moment $A_{LU}^{\sin\phi}$ is proportional to the polarized structure function $\sigma_{LT^\prime}$ \cite{Dre1992}:
\begin{equation}\label{eq:ALU}
	A_{LU}^{\sin\phi} = \frac{\sqrt{2 \epsilon (1 - \epsilon)}~\sigma_{LT^{\prime}}}{\sigma_{T} + \epsilon \sigma_{L}},
\end{equation}
where the structure functions $\sigma_{L}$ and $\sigma_{T}$ correspond to longitudinal and transverse virtual photons, and $\epsilon$ describes the ratio of their fluxes.

Hard exclusive $\pi^+$ electroproduction was measured at Jefferson Lab with the CEBAF Large Acceptance Spectrometer (CLAS) \cite{BM03}. Beam-spin asymmetries were extracted over a wide range in $Q^2$, $t$, $x_{B}$ and $\phi$. The incident electron beam was longitudinally polarized and had an energy of 5.498~GeV. The target was unpolarized liquid hydrogen. The CLAS detector consisted of six identical sectors within a toroidal magnetic field. The momentum and the charge polarity of the particles were determined by 3 regions of drift chambers from the curvature of the particle trajectories in the magnetic field. The electron identification was based on a lead-scintillator electromagnetic sampling calorimeter in combination with a Cherenkov counter. For the selection of deeply inelastic scattered electrons, cuts on $Q^{2}$~$>$~1~GeV$^{2}$ and on the invariant mass of the hadronic final state $W$~$>$~2~GeV were applied. Positive pions were identified by time-of-flight measurements. To select the exclusive $e^{\prime} \pi^{+} n$ final state, events with exactly one electron and one $\pi^{+}$ were detected, and a cut around the neutron peak in the missing mass spectrum was performed. The mean signal-to-background ratio in the forward region is 15.3, while it decreases to 4.9 in the backward region. 

Beam-spin asymmetries (BSA) were measured in the $Q^{2}$ range from 1 to 4.6 GeV$^2$, $-t$ up to 6.6 GeV$^2$ and $x_{B}$ from 0.1 - 0.6. The BSA and its statistical uncertainty were determined experimentally from the number of counts with positive and negative helicity ($N^{\pm}_{i}$), in a specific bin $i$ as:
\begin{eqnarray}
	BSA = \frac{1}{P_{b}} \frac{N^{+}_{i} - N^{-}_{i}}{N^{+}_{i} + N^{-}_{i}}~,
	~\sigma_{BSA} = \frac{2}{P_{b}} \sqrt{\frac{N^{+}_{i}  N^{-}_{i}}{(N^{+}_{i} + N^{-}_{i})^{3}}},~
\end{eqnarray}
\newline where $P_{b}$ is the average magnitude of the beam polarization. $P_{b}$ was measured with a M{\o}ller polarimeter upstream of CLAS and was 74.9 $\pm$ 2.4\%~(stat.) $\pm$ 3.0\%~(sys.).

\begin{figure}[h!]
	\centering
		\includegraphics[width=87mm,height=40mm]{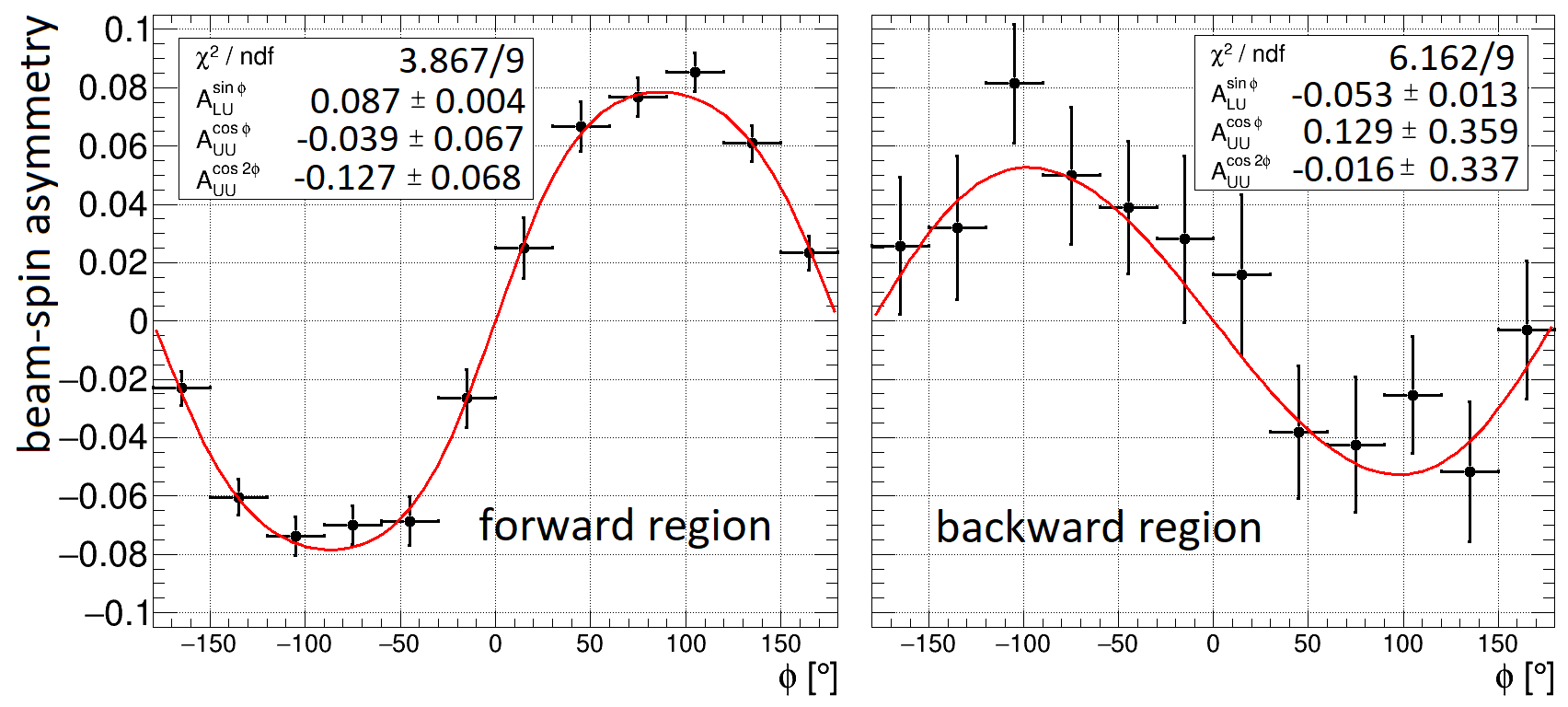}
	\caption{Beam-spin asymmetry as a function of $\phi$ for $\pi^{+}$ emitted in the forward (left) and backward (right) regions, integrated over all other kinematic variables. The vertical error bars show the statistical uncertainty of each point, while the horizontal bars correspond to the bin width. The red line shows the fit with the functional form of Eq. (\ref{eq:BSA}).}
	\label{fig:sinphi_fit}
\end{figure}

To extract the $\sin\phi$ moment $A_{LU}^{\sin\phi}$, the beam-spin asymmetry was measured as a function of the azimuthal angle $\phi$. Then a fit of the data with the functional form shown in Eq.~(\ref{eq:BSA}) was applied. Figure \ref{fig:sinphi_fit} shows the beam-spin asymmetry as a function of $\phi$ for events in the forward and backward regions, integrated over all other kinematic variables. Experimentally the forward region is defined as $\cos\theta_{CM}$~$>$~0 and $-t$~$<$~1.5~GeV$^{2}$, while the backward region is defined by a cut on $\cos\theta_{CM}$~$<$~0 and $-u$~$<$~2.0~GeV$^{2}$, where $\theta_{CM}$ is the polar angle of the pion in the frame boosted along the momentum transfer $\vec{q}$ direction. 
As expected the $\phi$-dependence can be well described by Eq. (\ref{eq:BSA}). The asymmetry of the background has been extracted with the side-band method by selecting a missing-mass interval on the right side of the missing neutron peak. These events represent the background under the region of interest and therefore its asymmetry has to be subtracted. The amplitude of the background asymmetry has been determined in the same way as for the exclusive events, with a $\sin\phi$ fit of the $\phi$-dependence of the BSA. The $\sin\phi$ amplitude of the background is $0.032 \pm 0.006$ in the forward region and decreases to $0.00 \pm 0.01$ in the backward region. Based on the signal-to-background ratio determined from a fit of the missing mass spectrum in each kinematic bin, a bin-by-bin background subtraction has been performed for the extracted $A_{LU}^{\sin\phi}$ values.

Several sources of systematic uncertainty were investigated, including particle identification, background subtraction, beam polarization, and the influence of the $A_{UU}^{\cos\phi}$ and $A_{UU}^{\cos2\phi}$ moments. The correlation between the unpolarized moments and $A_{LU}^{\sin\phi}$ was found to be very small. The systematic uncertainty for each contribution was determined by a variation of the contributing source around its nominal value. To estimate the impact of acceptance effects, a Monte Carlo simulation which included a parametrization of the kinematic behaviour following that of the actual data was performed. The impact of acceptance effects turned out to be small and is included in the systematic uncertainty. The total systematic uncertainty in each bin is defined as the square-root of the quadratic sum of the uncertainties from all sources. It has been found to be comparable to the statistical uncertainty.

\begin{figure}[h]
	\centering
		\includegraphics[width=70mm,height=55mm]{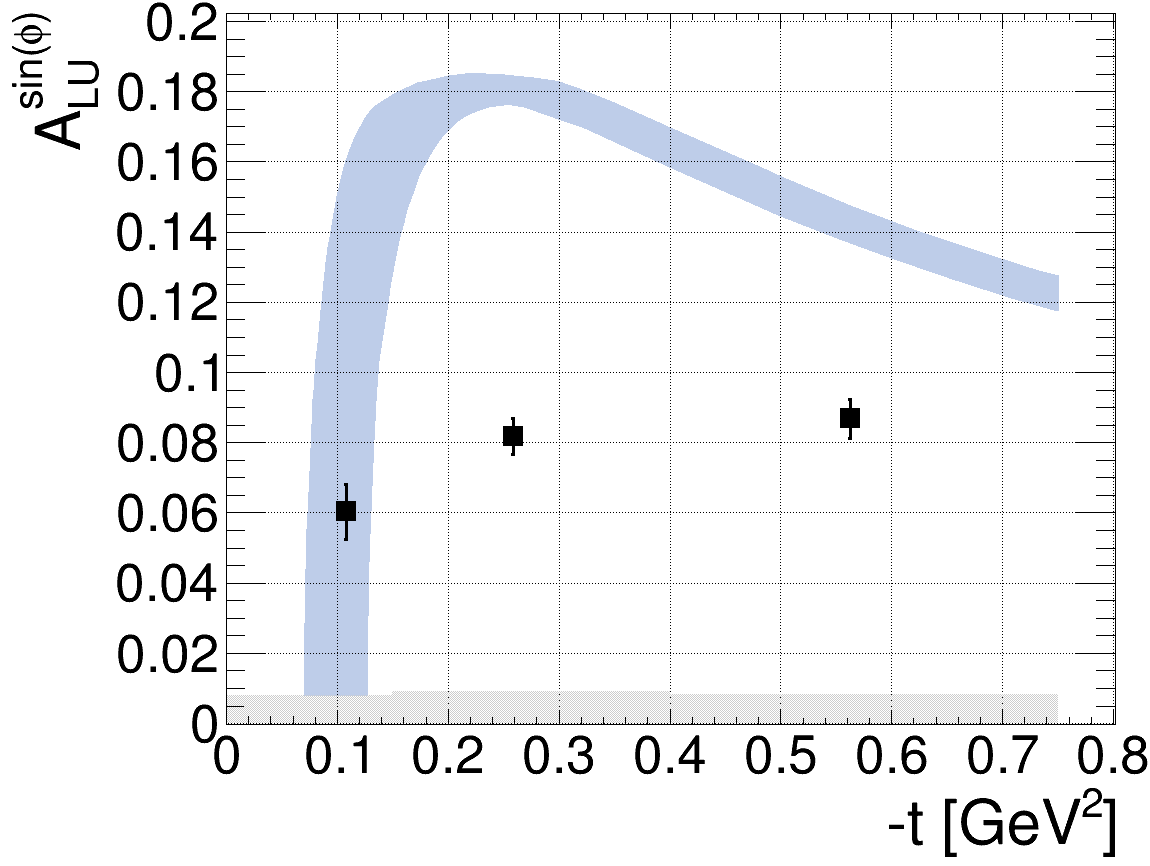}
	\caption{$A_{LU}^{\sin\phi}$ (black rectangles) as a function of $-t$ in the forward kinematic regime and their systematic uncertainty (grey bins). For comparison the theoretical prediction from the GPD-based Goloskokov-Kroll model (blue band) is shown. The band of the theoretical prediction corresponds to the range accessible with our measurements in $Q^{2}$ and $x_{B}$.}
	\label{fig:ALU_theory}
\end{figure}

Figure \ref{fig:ALU_theory} shows the results for $A_{LU}^{\sin\phi}$ in the region of $-t$ up to 0.75~GeV$^{2}$ ($-t/Q^{2} \approx 0.25$) where the leading-twist GPD framework is applicable and compares them to the theoretical predictions from the GPD-based model by Goloskokov and Kroll (GK) \cite{GK09}. The experimental data is binned in $-t$ and integrated over the 
complete $Q^{2}$ distribution ranging from 1.0 to 4.5~GeV$^{2}$ and $x_{B}$ ranging from 0.1 to 0.6. The band on the theoretical prediction represents the range in $Q^{2}$ and $x_{B}$ accessible with our measurements.
The GK model includes chiral-odd GPDs to calculate the contributions from the transversely polarized virtual photon amplitudes, with their $t$-dependence incorporated from Regge phenomenology. The GPDs are constructed from double distributions and constrained by results from lattice QCD and transversity parton distribution functions \cite{GK09}. A special emphasis is given to the GPDs $H_{T}$ and $\overline{E}_{T} = 2 \widetilde{H}_T + E_T$, while contributions from other chiral-odd GPDs are neglected in the calculations, unlike chiral-even GPDs, where some contributions are negligible but still included. The pion pole contribution to the amplitudes is taken into account for both the longitudinally and transversely polarized virtual photons. However, its contribution to the transversely polarized virtual photon amplitudes is very small. 

The magnitude of $A_{LU}^{\sin\phi}$ (see Eq. (\ref{eq:ALU})) is proportional to the ratio of the interference structure function $\sigma_{LT^\prime}$ and the unseparated cross section $\sigma_{0} = \sigma_{T} + \epsilon \sigma_{L}$, where $\sigma_{0}$ is forward peaked due to the pion pole term contribution and $\sigma_{LT^\prime}$ is constrained to be zero at $t=t_{min}~(\theta_{CM}=0)$ due to angular momentum conservation. 
$\sigma_{LT^\prime}$ can be expressed through the convolutions of GPDs with subprocess amplitudes (twist-2 for the longitudinal and twist-3 for the transverse amplitudes) and contains the products of chiral-odd and chiral-even terms \cite{previous1}:
\begin{eqnarray}
	\sigma_{LT^\prime} \sim Im\left[ \langle\overline{E}_{T-eff}\rangle^{*} \langle\widetilde{H}_{eff}\rangle + \langle H_{T-eff}\rangle^{*} \langle\widetilde{E}_{eff}\rangle\right],
\label{eqn:sigma_GPD}
\end{eqnarray}
where all involved GPDs are influenced directly or indirectly by the pion pole term, for example:
\begin{eqnarray}
	\widetilde{E}_{eff}~=~\widetilde{E}~+~pole~term,  \\
	\widetilde{H}_{eff}~=~\widetilde{H}~+~\frac{\xi^{2}}{1-\xi^{2}}~\widetilde{E}_{eff},
\end{eqnarray}
with the skewness $\xi \sim x_{B}/(2-x_{B})$. For $\pi^{+}$ the imaginary part of small chiral-odd GPDs in $\sigma_{LT^\prime}$ is significantly amplified by the pion pole term, which is real and exactly calculable. 
This feature increases the sensitivity of polarized observables to chiral-odd GPDs in contrast to the $\pi^{0}$ and $\eta$ channels where the pole contribution is
not present. In the GK model $\sigma_{LT^\prime}$ is dominated by $Im[\langle H_{T-eff}\rangle^{*}\langle\widetilde{E}_{eff}\rangle]$ and $\widetilde{E}_{eff}$ is dominated by the pion pole term, while other GPD products are considered to be negligible. 

The comparison between the experimental results and the theoretical predictions shows that the magnitude of the GK model calculations is overestimated and the $t$-dependence of the measured $A_{LU}^{\sin\phi}$ values shows a much flatter slope than the predicted curve. The discrepancy in magnitude and $t$-dependence might be due to the interplay of the pion pole term with the poorly known chiral-odd GPDs $H_{T}$ and $\overline{E}_{T}$. Even though previous studies showed that the GPD model can be well applied to predict $\pi^{0}$ and $\eta$ cross sections \cite{Bedlinskiy2012, Bedlinskiy2014, Bedlinskiy2017}, the results in Fig. \ref{fig:ALU_theory} show that the GPDs and the model have to be tuned to describe BSA as well.
While the beam-spin asymmetry calculations for the $\pi^+$ channel are overestimated by the GK model, the absence of the pion pole term in case of the $\pi^{0}$ and $\eta$ channels leads to a significantly smaller predicted beam-spin asymmetry by the GK model, which underestimates the experimental observation as shown in Ref. \cite{Zhao2019}. The combined analysis of these unique $\pi^{+}$ data with the $\pi^{0}$ and $\eta$ channels \cite{DeMasi2008, Zhao2019, Bedlinskiy2012, Bedlinskiy2014, Bedlinskiy2017} can be performed to significantly constrain these poorly known GPDs.

\begin{figure}[h!]
	\centering
		\includegraphics[width=70mm,height=55mm]{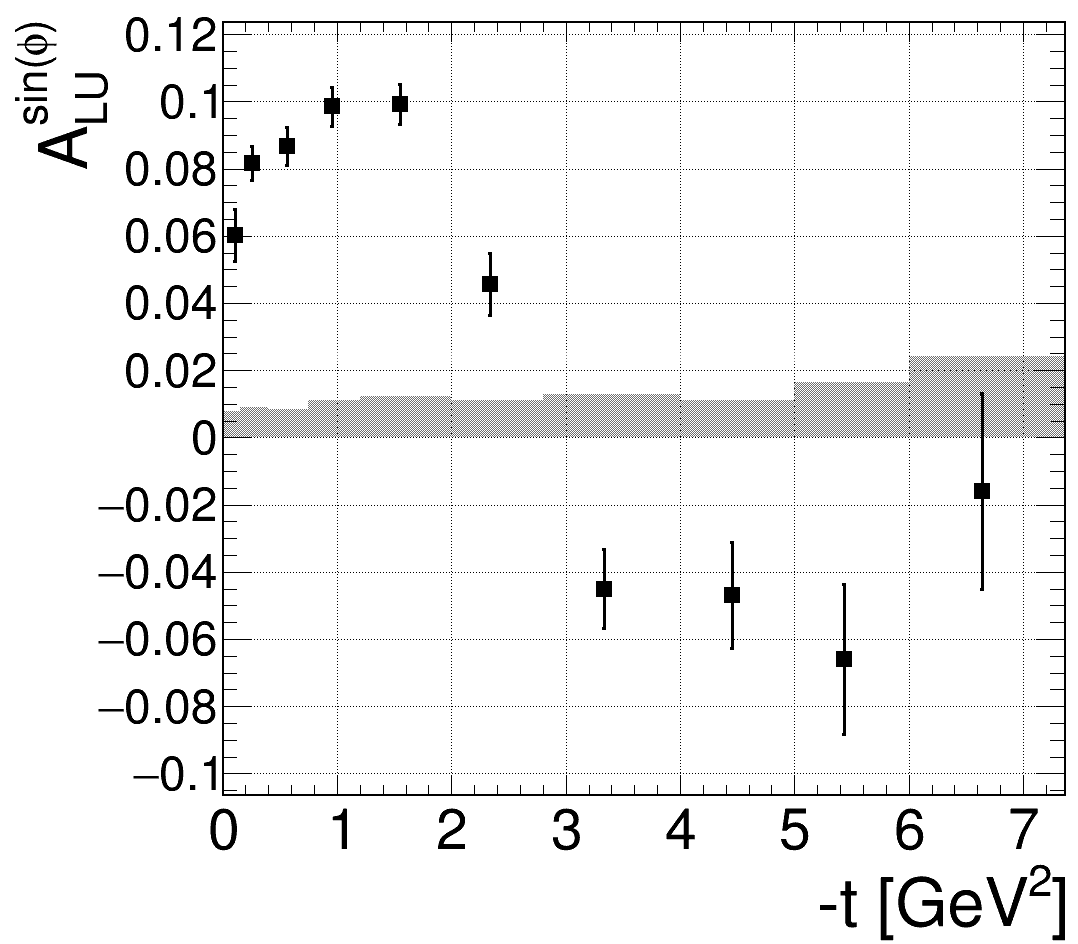}
	\caption{$A_{LU}^{\sin\phi}$ as function of $-t$. The shaded area represents the systematic uncertainty.}
	\label{fig:t-ALU}
\end{figure}

While the framework of GPDs is only applicable in very forward kinematics, a complete understanding of the reaction mechanism requires measurements over the complete range of $-t$. As shown in Fig. \ref{fig:t-ALU}, we extended the kinematic region for the extraction of $A_{LU}^{\sin\phi}$ up to $-t$~=~6.6~GeV$^{2}$, which is close to the maximal accessible $-t$ value. The data are binned in $-t$ and integrated over the complete $Q^{2}$ distribution ranging from 1~GeV$^{2}$ - 4.5~GeV$^{2}$ and $x_{B}$ ranging from 0.1 to 0.6. 

The sign of of $A_{LU}^{\sin\phi}$ in forward kinematics is clearly positive, which is confirmed by the most recent GPD models \cite{GK09}, while in backward kinematics a clearly negative sign is observed. Large $t$ corresponds to small $u$, so that at backward angles the u channel dominates (Fig. \ref{fig:production_mechanism} (b)). Thus it is expected that the TDA-based framework can be applied in very backward kinematics.

Similarly to Eq. (\ref{eqn:sigma_GPD}) for very forward kinematics, $\sigma_{LT^\prime}$ in the backward regime can be expressed through the interference between the leading twist transverse amplitude of the convolution in terms of twist-$3$ $\pi N$ TDAs ($H^{{\rm tw}3}$) and nucleon DAs ($\phi^{{\rm tw}3}$) and the next leading sub-process longitudinal amplitude of the convolution involving twist-$4$ TDAs ($H^{{\rm tw}4}$) and DAs ($\phi^{{\rm tw}4}$) \cite{PRD85, Braun:1999te, Braun:2000kw}:
\begin{eqnarray}
 \sigma_{LT^\prime}  \sim {\rm Im} \left[\langle H_i^{{\rm tw}3} \phi_j^{{\rm tw}3} \rangle \left( \langle  H_i^{{\rm tw}4} \phi_j^{{\rm tw}3} \rangle + \langle H_i^{{\rm tw}3} \phi_j^{{\rm tw}4} \rangle \right)^* \right].~
\end{eqnarray}
A complete theoretical study of this twist-$4$ longitudinal amplitude is not yet available, and it is an open question which particular twist-$4$ $\pi N$ TDAs and DAs will contribute to the BSA and what kind of phenomenological models can be implemented for these quantities. Nevertheless, our measurement will significantly constrain the nearly unknown TDAs and help to further develop the TDA-based framework.

Also, for the intermediate kinematic region around $\theta_{CM}$~=~90$^{\circ}$, first models have been introduced \cite{Diehl:2013, Kroll:2018}. However, calculations exist only for wide-angle Compton scattering \cite{Diehl:2013} and the photoproduction of pions \cite{Kroll:2018}. Nevertheless, the introduced concepts can also be applied to electroproduction and will help to connect the GPD and TDA kinematic regimes in the future.

As shown in Fig. \ref{fig:t-ALU}, the $t$-dependence of $A_{LU}^{\sin\phi}$ makes a clear transition from positive values with a maximum value of 0.10 in the forward region to negative values down to a minimum value of -0.06 in the backward region. The sign change occurs around $-t$~=~3~GeV$^{2}$, which corresponds to $\theta_{CM}$~=~90$^{\circ}$, and marks the transition between the $\pi^{+}$ emitted in the forward and backward directions. Therefore, the sign change may be interpreted as an indication for a transition between the GPD and TDA regimes. The wide range of kinematics presented in this work will also enable the development of a more consistent reaction mechanism for the intermediate kinematical regime in-between the very forward regime with GPD-based description and the very backward regime with description in terms of TDAs.

\begin{figure}[h!]
	\centering
		\includegraphics[width=85mm,height=40mm]{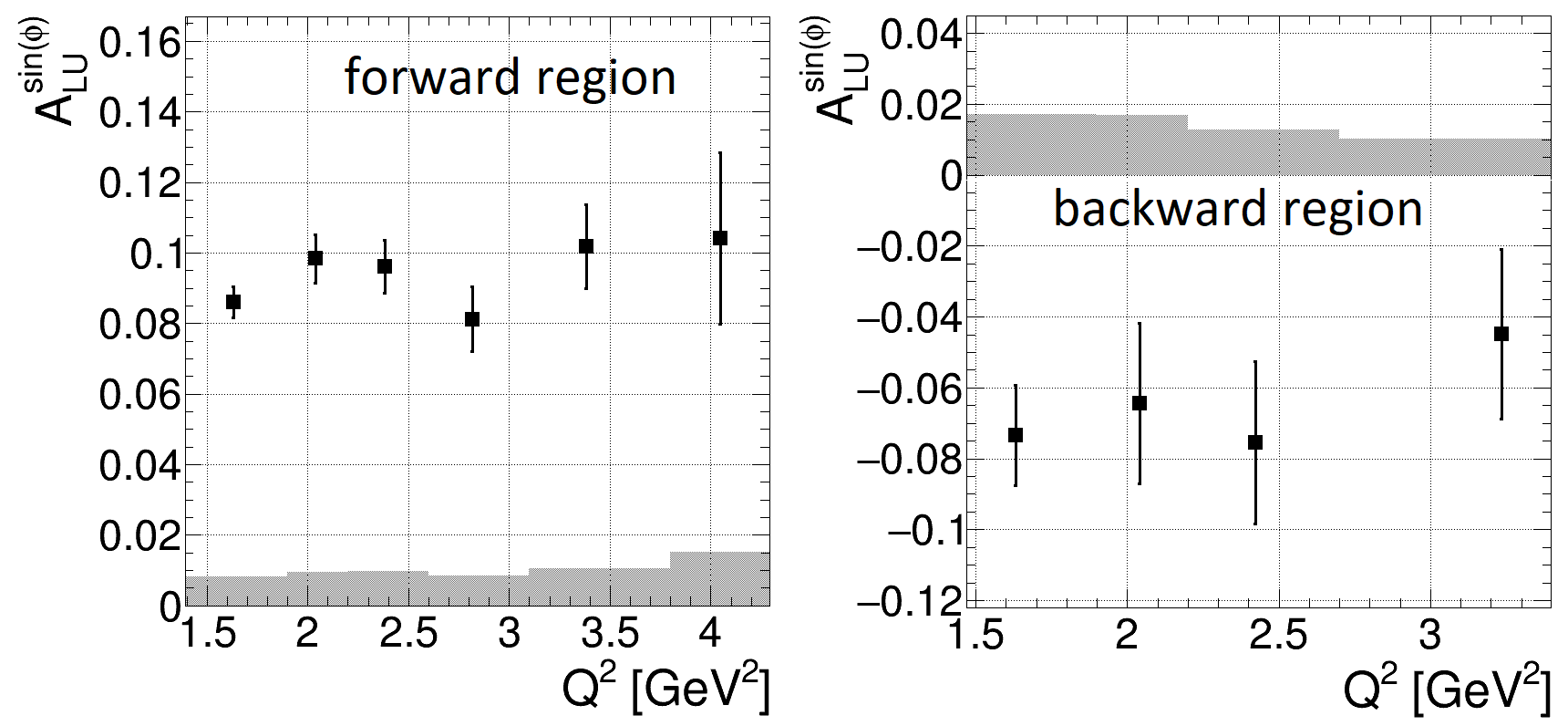}
		\includegraphics[width=85mm,height=40mm]{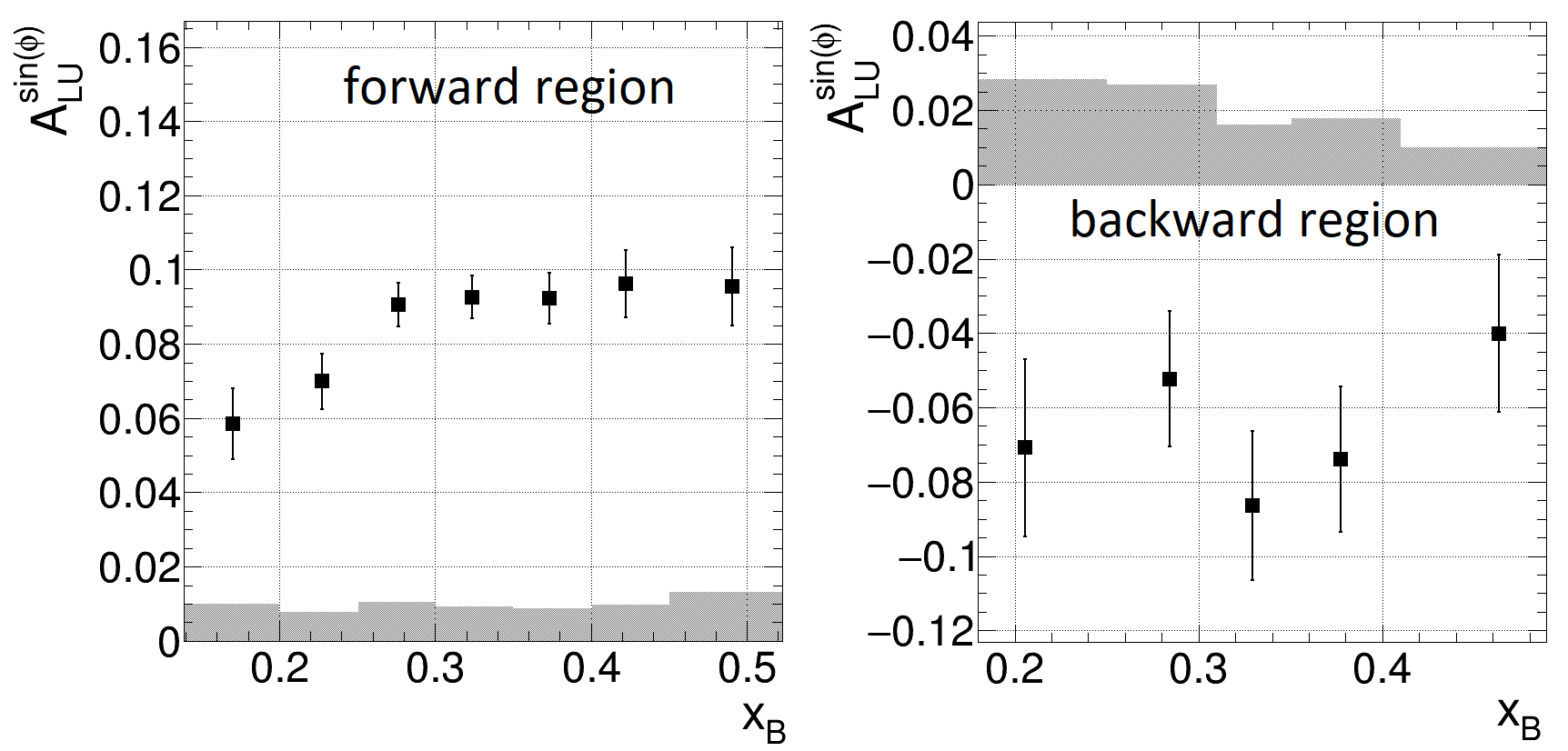}
	\caption{$A_{LU}^{\sin\phi}$ as function of $Q^2$ (top) and $x_B$ (bottom) for pions going in the forward (left) and backward (right) regions. The shaded area represents the systematic uncertainty.}
	\label{fig:q2_ALU}
\end{figure}

Figure \ref{fig:q2_ALU} shows $A_{LU}^{\sin\phi}$ as a function of $Q^{2}$, integrated over $x_B$ in the top plots and as a function of $x_{B}$, integrated over $Q^2$ in the bottom plots, for pions going in the forward (left) and backward (right) regions, as defined earlier.
The figure clearly shows that the sign change between the forward and the backward region is present for all $Q^{2}$- and $x_{B}$-bins.
In the forward region, the $Q^{2}$-dependence shows a rather flat behavior, while $A_{LU}^{\sin\phi}$ rises for small $x_{B}$ until it reaches a constant level for $x_{B}$~$>$~0.26. In the backward region the $Q^{2}$- and $x_{B}$-dependencies show a rather flat behavior. However, the effect is not statistically significant.

In summary, we have measured for the first time the $\sin\phi$ moment $A_{LU}^{\sin\phi}$ of beam-spin asymmetries for $\vec{e} p \to e^\prime n \pi^+$ at large photon virtuality, above the resonance region over the full range of polar angles $\theta_{CM}$ that cover the complete kinematic region of the GPD and TDA frameworks simultaneously. A comparison in very forward kinematics showed that our $A_{LU}^{\sin\phi}$ measurement cannot be described in magnitude or $t$-dependence by the most advanced GPD-based model \cite{GK09}. In very forward kinematics where the GPD framework is applicable, we measure clearly positive values of $A_{LU}^{\sin\phi}$, while in very backward kinematics where the TDA framework is applicable, negative $A_{LU}^{\sin\phi}$ values have been measured. A clear sign change of $A_{LU}^{\sin\phi}$ has been observed around $\theta_{CM} = 90^\circ$. The presented data provide important constraints for the development of a reaction mechanism that describes the complete kinematic regime including GPDs and TDAs as well as the intermediate regime. To obtain a deeper understanding, and to reveal more details of the reaction mechanism, measurements with a higher precision and over a larger range of $Q^2$ will be performed with the upgraded 12 GeV CEBAF accelerator at JLab and in the crossed reaction $\bar N N \to \gamma^* \pi$, accessible with \={P}ANDA at FAIR \cite{panda-papers1, panda-papers2, panda-papers3} and $\pi N \to N \gamma^*$ or $\pi N \to N J/\Psi$ at J-PARC \cite{JPARC16}. The data-set presented in this work can be downloaded from Ref. \cite{CLASdata}.


We acknowledge the outstanding efforts of the staff of the Accelerator and the Physics Divisions at Jefferson Lab in making this experiment possible.
We also acknowledge very helpful discussions with L. Szymanowski and B. Pire. This work was supported in part by the U.S. Department of Energy, the National Science Foundation (NSF), the Italian Istituto Nazionale di Fisica Nucleare (INFN), the French Centre National de la Recherche Scientifique (CNRS), the French Commissariat pour l$^{\prime}$Energie Atomique, the UK Science and Technology Facilities Council, and the National Research Foundation (NRF) of Korea. The Southeastern Universities Research Association (SURA) operates the Thomas Jefferson National Accelerator Facility for the US Department of Energy under Contract No. DE-AC05-06OR23177. The work is also supported in part by DOE grant no: DE-FG02-04ER41309.


\end{document}